\documentclass[preprint]{iucr}              
\usepackage{graphicx}
\usepackage{rotating}
\usepackage{longtable}
     \paperprodcode{a000000}      
     \paperref{xx9999}            
     \papertype{CP}               

     \paperlang{english}          
     \journalcode{J}              
     \journalyr{2007}
     \journaliss{1}
     \journalvol{63}
     \journalfirstpage{000}
     \journallastpage{000}
     \journalreceived{0 XXXXXXX 0000}
     \journalaccepted{0 XXXXXXX 0000}
     \journalonline{0 XXXXXXX 0000}

\newcommand{\s}{{section}} 

\begin{document}                  

\title{McSAS: A package for extracting quantitative form-free distributions}
\shorttitle{Nanoscale quantification with McSAS}

\author[a]{I.}{Bre\ss ler}{ingo.bressler@bam.de}
\author[b]{B. R.}{Pauw}{brian@stack.nl}
\author[a]{A.}{Th\"unemann}

\aff[a]{BAM Federal Institute for Materials Research and Testing, 12205 Berlin, Germany}
\aff[b]{National Institute of Materials Science, 305-0047 Tsukuba, Japan}

\shortauthor{Bre\ss ler, Pauw, and Th\"unemann}

\maketitle



%

\begin{abstract}

A reliable and user-friendly characterisation of nano-objects in a target material is presented here in the form of a software data analysis package for interpreting small-angle X-ray scattering (SAXS) patterns. 
When provided with data on absolute scale with reasonable uncertainty estimates, the software outputs (size) distributions in absolute volume fractions complete with uncertainty estimates and minimum evidence limits, and outputs all distribution modes of a user definable range of one or more model parameters. 
A multitude of models are included, including prolate and oblate nanoparticles, core-shell objects, polymer models (Gaussian chain and Kholodenko worm) and a model for densely packed spheres (using the LMA-PY approximations). 
The McSAS software can furthermore be integrated as part of an automated reduction and analysis procedure in laboratory instruments or at synchrotron beamlines. 

\end{abstract}


\section{Introduction}

Quantification of nanoscale structures is set to become a requirement in industrial preparation of materials \cite{Commission-2011}. 
An accurate and reliable toolset is therefore sought to obtain quantitative size distributions of disperse nanoparticle mixtures, spanning a wide size-range with minimal effort and high flexibility. 

The most commonly used technique for nanostructural quantification is transmission electron microscopy (TEM). 
TEM is essential in determining the overall morphology of the nanostructural features, and can often be used to coarsely quantify their parameters. 
Obtaining a statistically representative quantification of the nanostructure, however, is reliant on the probing of large numbers of objects. 
To improve its representation of the bulk of the sample, it should preferably be performed through sampling from multiple locations throughout a bulk-scale sample \cite{Klein-2011,Meli-2012}.
As TEM has remained largely resilient to automation efforts, this commonly remains a tedious and labour-intensive task. 
It is therefore beneficial to combine the localised resolving power of microscopy with another technique more suited for bulk-scale nanostructural quantification such as Small-angle Scattering (SAS) \cite{ISO-2014,Pauw-2013a}. 

SAS offers one reliable route to bulk quantification of materials: it can characterise the nanostructure of large amounts of material with a minimum of tedium, easily extracting size distributions and volume fractions. 
Practically, however, one of the biggest stumbling blocks in its application has been the data correction and analysis, in particular for the disperse systems discussed herein. 
While the discussion of data corrections is beyond the scope of this work (c.f. \cite{Jacques-2012,Pauw-2013a,Kieffer-2013}), it has to be stressed that reliable analysis of data is reliant on the quality thereof. 
There can be no good results without proper data which cannot be considered complete without reasonable uncertainty estimates.

Analysis can be performed through a ``classical'' approach: by least-squares optimisation of model function parameters describing a scattering pattern \cite{Pedersen-1997}.
However, the assumptions made in such model functions (on the scatterer shape and its size distribution) are often insufficiently flexible to describe the morphology of many samples. 
Good agreement between the model function and the measured data will then not be achieved, in particular for samples where the actual size dispersity does not adhere to the inherently assumed model size distribution shape. 

Modern analysis methods are available for this class of samples (size-disperse) which allow for the retrieval of size distributions without assumptions on its form. 
While the general shape of the scatterer still has to be defined in order to arrive at a unique solution (see, for example \cite{Rosalie-2014}), these modern methods are no longer restricted to a limited set of idealized size distributions.

There are a variety of approaches for these modern methods, including indirect fourier transform methods based on smoothness criteria \cite{Glatter-1977,Svergun-1991}, maximum entropy optimisation \cite{Hansen-1991} or Bayesian hyperparameter estimation \cite{Hansen-2000}. 
Alternative methods include using Titchmarsh transforms for extraction of size distributions \cite{Fedorova-1978,Botet-2012}. 
While these carry a certain mathematical elegance, they can be challenging to implement, understand and apply. 
This mathematical obscurity furthermore hinders thorough understanding of the failure modes, which can lead to crucial errors in their application. 

Recently, a (conceptually) straightforward method was presented for determining size distributions from small-angle scattering patterns \cite{Pauw-2013}, which has since been applied to a variety of sample types including metal alloys \cite{Rosalie-2014}, novel oxygen reduction reaction catalysts \cite{Schnepp-2013} and polymer fibres \cite{Pauw-A2013}. 
While these results have been encouraging, the lack of user-friendliness of the method has hindered its adoption by a broader audience. 

Through a multinational, collaborative effort spanning several years, a drastic improvement on the software usability was effected. 
This was mostly accomplished through a comprehensive rewrite of the implementation following modern coding standards and conventions, and the addition of a graphical user interface (c.f. Figure \ref{fg:main}). 
After a brief recapitulation of the method concept, the software capabilities and interface are detailed, and an application example is given.

\section{McSAS fitting procedure}

\subsection{Concept overview}

\begin{figure}
	\begin{center}
		\includegraphics[width=0.95\textwidth]{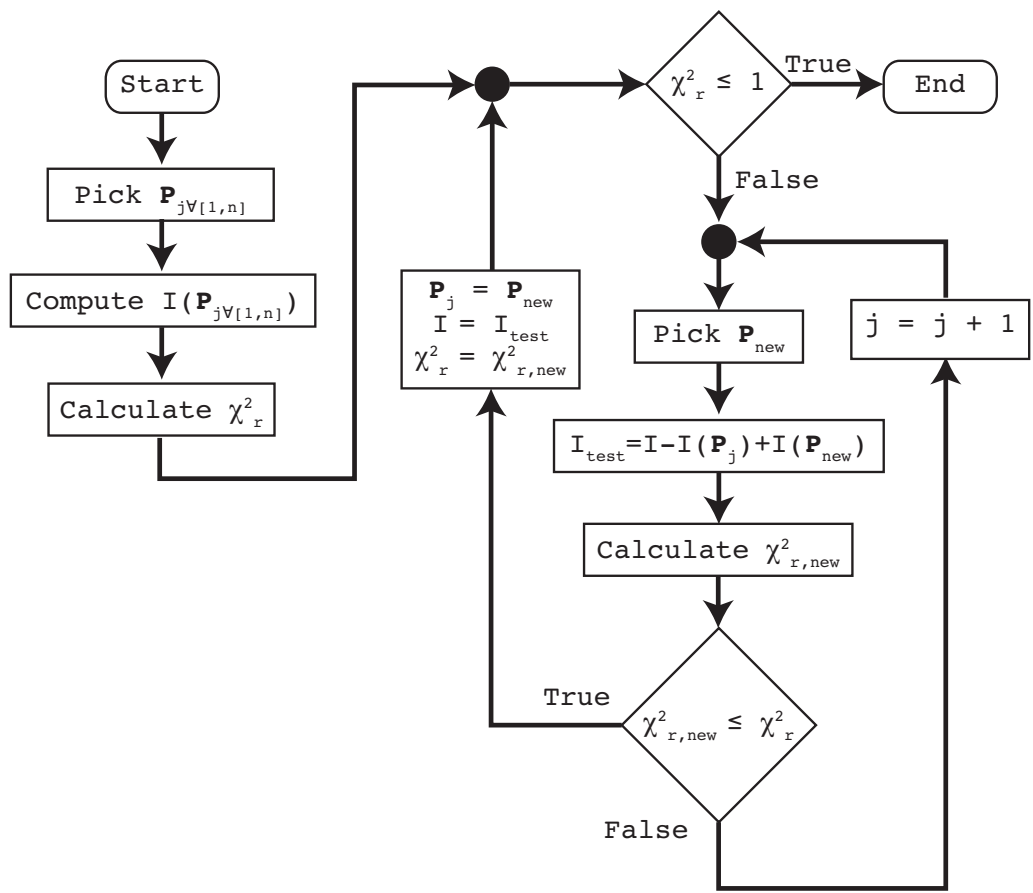}
		\caption{The main process of the McSAS software for parameter optimisation.} \label{fg:process}
	\end{center}
\end{figure}


The McSAS method is a Monte Carlo Rejection Sampling approach for retrieving size distributions from scattering patterns \cite{Pauw-2013}. 
The core procedure is shown in Figure \ref{fg:process}. 
The method starts from a set of non-interacting nanoparticles of predefined shape (e.g. spheres, rods, ellipsoids) but with random values chosen for the active fitting parameter(s). 

The scattering effected by this set is computed as the sum of the contributions. 
During the optimisation procedure, the scattering pattern of the contributions are (typically) weighted with the inverse of their volume to obtain a more representative impact on the total. 
This process drastically reduces the required number of contributions and improves fitting flexibility\protect\footnote{Thus, about 200 contributions are typically sufficient to describe the majority of scattering patterns with a minimum of optimisation iterations}. 
This weighting is reversed at the end of the optimisation. 

A figure of merit ($\chi^2_r$) is calculated from the model based on its agreement with the measured data, with due consideration of the data uncertainty estimates \cite{Pedersen-1997}. 
In order to obtain this figure of merit, the model intensity is matched to the measured dataset through scaling and addition of an optional flat background contribution.
The scaling and background parameters are obtained by least-squares minimisation of $\chi^2_r$.
The figure of merit thus indicates when the model describes the data on average to within the data uncertainty ($\chi^2_r \leq 1$). 
A suitable cut-off criterion for optimisation is thereby provided that generally prevents over-fitting of the data, and allows for the estimation of uncertainties on the resultant distribution.

Each iteration of the MC procedure consists of replacing one of the scattering objects in the set by another object of the same basic shape but different, randomly chosen values for its active fitting parameter(s). 
This replacement is accepted if it reduces $\chi^2_r$, i.e. if the agreement of the resulting MC scattering pattern with the measured pattern is improved. 
These iterations continue until the convergence criterion of $\chi^2_r \leq 1$ is reached (different convergence criteria can be set by the user for data whose uncertainties are over- or underestimated).

The size distribution is determined through analysis of the active fitting parameter values throughout the set. 
At this stage, the scattering contrast is taken into account to establish the absolute volume fractions. 
Finally, the uncertainty on the size distribution is determined through analysis of the sample standard deviation of a multitude of independent MC solutions. 

In addition to this, a ``minimum observability limit'' is defined for each size, which specifies the minimum required volume fraction of scatterers required to make a \emph{distinguishable} contribution to the scattering pattern (i.e. a contribution exceeding the measurement uncertainty).
More specifically, a minimum observability limit $\varphi_{\mathrm{min},k}$ (in units of volume fraction) can be defined for any method where the total model intensity is comprised of a set of quantised components, whose partial contributions are $I_k(q)$ for a given component volume fraction $\varphi_k$, and where the measurement data uncertainty $\sigma (q)$ is available:

\begin{equation}\label{eq:phimino}
\varphi_{\mathrm{min},k}= \min\limits_{q \in [q_{\mathrm{min}},q_{\mathrm{max}}]}\left[ \frac{\sigma(q)\varphi_k}{I_k(q)} \right]
\end{equation}

Its derivation and use is further explored elsewhere \cite{Pauw-2013}.

The uncertainty estimates and the observability limits are key values in the application of the method. 
They provide information to distinguish between numerical noise and size distribution components which are evidenced by the data, and furthermore allow for the assessment of the statistical significance of differences in resultant size distributions. 
The accuracy of the uncertainty estimates and observability limits are, however, directly reliant on the provision of reasonable uncertainty estimates on the measured data.

Using the aforementioned method, McSAS is able to retrieve any freeform size distribution provided a basic scatterer shape is given. 
A test of the retrievability of a wide range of unimodal and multimodal size distributions has been demonstrated for a large variety of simulated size distributions in the supplementary information of Pauw et al. \cite{Pauw-2013}. 
A comparison between size distributions of precipitates in alloys is also available, obtained from electron microscopy and McSAS analysis of SAXS data \cite{Rosalie-2014}.

\subsection{MC method benefits and drawbacks}\label{sssc:drawbacks}
The MC method has already demonstrated its versatility in a variety of practical applications, including monitoring of precipitation processes in alloys and bulk validation of nanoparticle sizes in catalysts. 
It can work in absolute units to retrieve absolute volume fractions. 
Several scatterer shapes have been implemented, and a model for densely packed spheres has also been included based on the Local Monodisperse Approximation (LMA)\protect{\footnote{This is one of the few structure factors that can be directly implemented given the internal design of the MC method.}} coupled with the Percus-Yevick (PY) approximation (example given \textit{vide infra}) \cite{Kinning-1984}. 

On the downside, due to its ``brute force'', iterative nature, the method is not as fast as some of the alternatives. 
Optimisation of a reasonably accurate dataset can take a minute or more on a modern desktop computer. 
This is expected to improve in the near future through implementation of multithreading.

Furthermore, there is a risk of under-specifying the fitting model when more complex models are chosen. 
For example, if a cylindrical scatterer model is chosen, and its length and radius allowed to span the same size range, the solution is no longer unique and a multitude of valid solutions will be found. 
This is indicated by excessive uncertainties resulting from large discrepancies between the independent McSAS repetitions, and such ambiguity can be easily arrived at when using models such as core-shell objects and anisotropic objects. 
For such shapes, the allowed size ranges for the shape parameters may require the application of strict limits before a unique solution is obtained. 

Lastly, experimentally obtained SAS data represents a volume-weighted size distribution for broad size distributions rather than a number-weighted size distribution due to the physical weighting of contributions in SAS experiments. 
In other words, there is not a lot of evidence in SAS data to allow the extraction of broad number-weighted size distributions (explored, amongst others, in \s\,3 of \cite{Pauw-2013}). 
For this reason, when trying to extract number-weighted size distributions using McSAS, excessively large contributions, uncertainties and observability limits are often seen at small sizes. 
It is therefore recommended to extract volume-weighted size distributions with McSAS instead. 
While this is not strictly a drawback of the McSAS method (but rather a limitation of SAS measurements on disperse systems), it stands in strong contrast with classical methods. 
In classical methods, the strict assumptions placed on the size distribution shape belie the lack of evidence for the absence or presence of scatterers at the small end of the distribution (and the number-weighted distributions are mostly determined by the evidence for the larger sizes alone).


\section{Current implementation}

\subsection{Feature overview}
Besides the sphere model, a variety of alternative models are also provided. 
Special attention should be given to prevent fitting ambiguity for these more complex shapes. 
These include core-shell ellipsoids and core-shell spheres, a Kholodenko worm and Gaussian chain model, isotropic cylinders and a sphere model suitable for dense packings of size-disperse spheres.

The code is unit-aware, and can translate between the units used in the user interface (often chosen based on tradition, such as \AA$^{-2}$ for the scattering length density difference and $^\circ$ for angle), and the SI units used for the internal calculations (m$^{-2}$ and radian, respectively). 
In our experience, the strict internal adherence to SI units drastically reduces the risk of programming errors.

Once set up, the MC code can run with or without user interface, to allow integration into existing data processing procedures. 
When run using the interface, the data file(s) can be loaded from the command line, if necessary using regular expressions. 
The fitting procedure can then be automatically initiated, inheriting the settings of the previous GUI instance.

To aid the user with the provision requirement of reasonable uncertainty estimates, any provided uncertainties will be limited to be no less than 1\% of the magnitude of the scattering signal. 
This has been previously found to be a practical limit from data correction considerations \cite{Pauw-2013}, and is a value supported by experimental results \cite{Hura-2000}. 
Naturally, this lower limit can be adapted or bypassed if better estimates can be guaranteed.

Graphical output and population statistics can be calculated for a user-specified number of parameter ranges, with the size axes in logarithmic or linear scales. 
The distributions can be shown as number-weighted or volume-weighted size distributions, though for broad distributions volume-weighting is strongly recommended (see the note in \s\,\ref{sssc:drawbacks}). 
The size distributions shown include the observability limit, i.e. the minimum required amount for each contribution to be statistically significantly contributing. 

Lastly, population statistics of the solution are determined independently of the histogramming procedure. 
For each selected parameter and range, the total value and the four distribution modes are provided: the mean, variance, skew and kurtosis. 
These are number- or volume-weighted depending on the user's choice. 
Such parameters simplify the analysis of population trends in in-situ experiments or other inter-related datasets.

\subsection{User interface features}

\begin{figure}
	\begin{center}
		\includegraphics[width=0.85\textwidth]{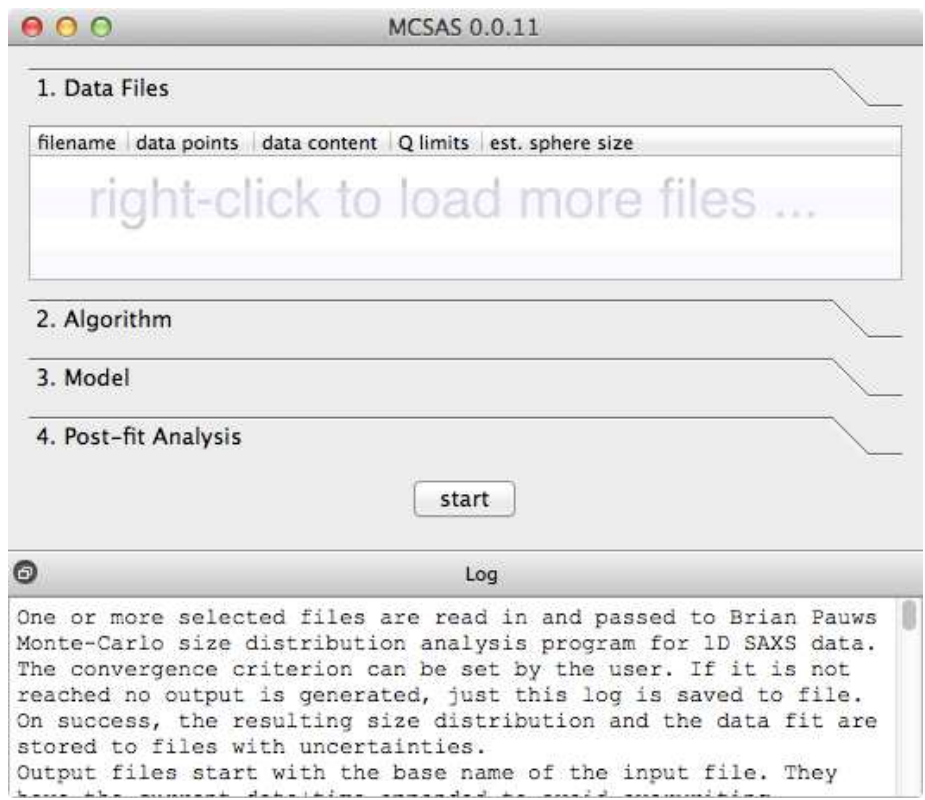}
		\caption{The main interface of the McSAS software upon start-up, showing four configuration panels. The "Data Files" panel allows selection and input of the data of interest, the "Algorithm" panel contains settings to adjust the optimisation method behaviour, "Model" contains all parameters and settings relevant to the chosen morphology, and "Post-fit Analysis" holds the settings for histogramming and visualisation of the result.} \label{fg:main}
	\end{center}
\end{figure}

The user interface is divided into several panels, each limited to a different aspect of the process (See Figure \ref{fg:main}). 
These consist of a ``Data files''-panel, an ``Algorithm''-panel, a ``Model''-panel and a ``Post-fit Analysis''-panel, and will be discussed forthwith.

The ``Data files''-panel shows datafiles loaded upon startup (as command-line arguments) or files added through the right-click menu. 
All files will be treated identically when the fit is run, though their order of processing can be changed as desired. 
Available data is guessed from the input file, which is expected to consist of three semi-colon separated columns of $q$\,(nm), $I$\,((m sr)$^{-1}$) and the uncertainty estimate $\sigma(I)$\,((m sr)$^{-1}$). 
An optional fourth column can be used to indicate the azimuthal angle $\phi$ to aid fitting of anisotropic scattering patterns. 

To aid the user with determining reasonable limits of size parameters, basic analysis is performed when loading each data file. 
The minimum and maximum values of the provided $q$-vector are used to estimate the maximum and minimum possible scatterer size under the assumption of solid spherical scatterers. 
Those estimates are displayed next to each data file and, by double-click, can be applied as optimisation limits for radius-type model parameters.

The ``Algorithm''-panel contains a subset of MC algorithm settings addressable by the user. 
The most important of these is the chi-squared criterion. 
While this is per default set to 1, it may prevent reaching a state of convergence ($\chi^2_r \leq 1$) with insufficiently large or poorly estimated uncertainty values. 
Increasing this value will allow the convergence condition to be reached, after which the fit may be evaluated. 
This increase directly affects the uncertainties on the resultant distribution.

Additionally, the number of shape contributions can be increased. 
While the default setting of 200 is sufficiently large to reach the convergence criterion for most scattering patterns, and small enough to reach it rapidly, there may be cases for which an increased number is desired. 
Using the timing information shown in the graphical output, the number of shape contributions can be optimised to reach convergence as fast as possible (a method discussed in \cite{Pauw-2013}). 
Likewise, the number of repetitions can be changed. 
These independent repetitions are used to estimate the uncertainties on the resultant size distribution, but a reduced number will suffice for initial testing. 
Here too, a selection can be made on whether a flat background contribution is to be taken into account when matching the MC intensity to the detected signal. 

The ``Model''-panel contains all information on the model used to describe the scatterer morphology. 
The pulldown menu offers a selection of models that can be used as basic scatterer. 
The associated parameters and options for the model chosen will then be shown on the right-hand side. 
The models allow setting of a scattering length density difference as well, which (when used in combination with absolute units in the input intensity) will result in absolute volume fractions in the final distribution. 
Some parameters can be set ``Active'', which implies they are active fitting parameters for the MC model. 
When set, lower and upper limits have to be provided for the parameter to vary between.

The ``Post-fit Analysis''-panel offers basic analysis capabilities for interpretation of the MC result. 
When a range entry is added, the user can select which parameter to histogram, what parameter range to consider, and how many bins to histogram in. 
Increasing the number of histogram bins will lead to increased detail in the resulting histogram at the cost of larger uncertainties and evidence requirements \textit{via} observability limits. 
Furthermore, a choice can be made whether to use a linear or logarithmic parameter-scale (useful for distributions spanning several decades) and whether to plot volume- or number-weighted size distributions. 
When using absolute units, only the volume-weighted distribution will contain absolute values, the number-weighted distribution is normalised for lack of information. 
Again it should be stressed that the evidence in SAS experiments supports volume-weighted rather than number-weighted distributions (\emph{vide supra}). 

Finally, the ``Start''-button starts the process, and the "log" shows the output of the program as it runs (and is automatically stored in a file). 


\subsection{Support, availability and licensing}
A reasonable degree of support is provided by the authors subject to the availability of time. Instructional videos are available to help the user get started. 

The software is written in the Python 2.7 programming language, and available as a Git DVCSS repository. 
Binary packages of stable versions are available for selected operating systems. 
The software is released under an open-source GPLv3 license, allowing for academic and commercial adoption given proper attribution. 
Users for whom the software has been useful may refer to this work. 

\section{Application example to densely packed nanoparticles}

Dense systems add a degree of complexity to small-angle scattering, and are therefore interesting as a test case for MC methods. 

A suitable dataset of densely packed, dry SiO$_2$ spheres (with a stated radius of 75\,nm) has been provided by Peter H\o gh\o j of Xenocs, as part of a demonstration dataset measured on their Xeuss SAXS instrument. 
The SiO$_2$ spheres are packed in a randomly jammed fashion, implying that the volume fraction $v_f$ is approximately 0.63 \cite{Song-2008}. 


\begin{figure}
	\begin{center}
		\includegraphics[width=0.85\textwidth]{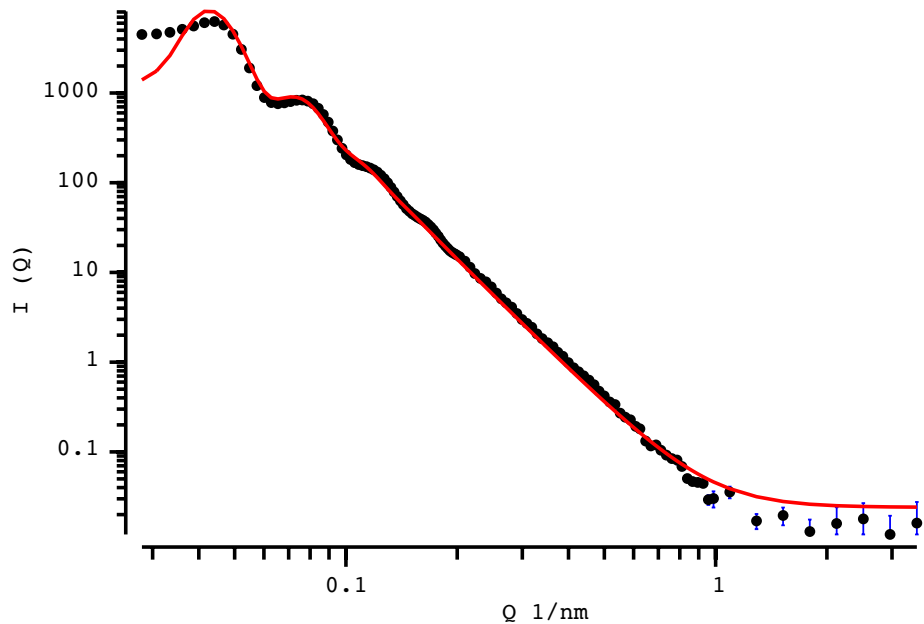}
		\caption{Best fit using a classical model (implemented in SASFit) to a scattering pattern obtained from packed SiO2 spheres. Model uses a sphere form factor with a LMA-PY structure factor and a Gaussian size distribution.} \label{fg:densesasfit} 
	\end{center}
\end{figure}


A reasonable fit can be obtained using classical fitting methods implemented in SASfit, with a model of Gaussian distributed spheres and a structure factor consisting of a PY hard-sphere interaction model assuming the local monodisperse approximation (LMA). 
This resulting fit is shown in Figure \ref{fg:densesasfit}. 
Most of the intensity can be described well, apart from the region at low $q$, which is remarkable given that the structure factors are not strictly valid for application to very dense systems. 

\begin{figure}
	\begin{center}
		\includegraphics[width=0.85\textwidth]{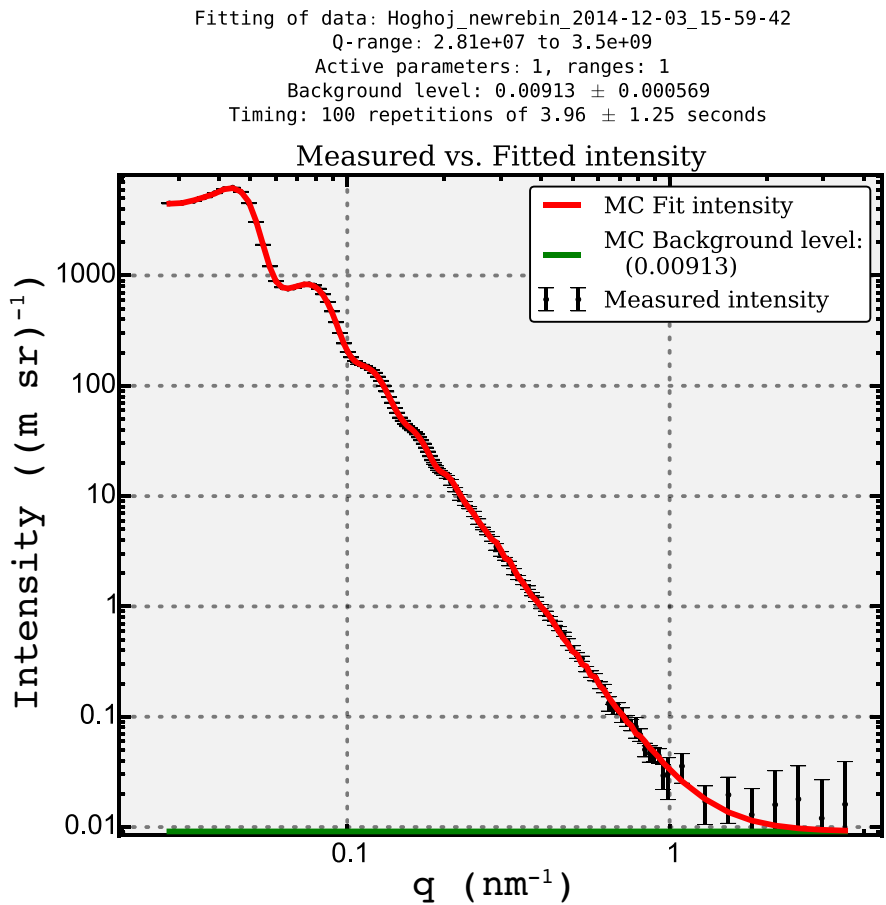}
		\caption{McSAS graphical output panel showing the best fit obtained using the MC method to a scattering pattern obtained from packed SiO2 spheres. 
		Model using a sphere form factor with a LMA-PY structure factor.} \label{fg:denseI} 
	\end{center}
\end{figure}

\begin{figure}
	\begin{center}
		\includegraphics[width=0.85\textwidth]{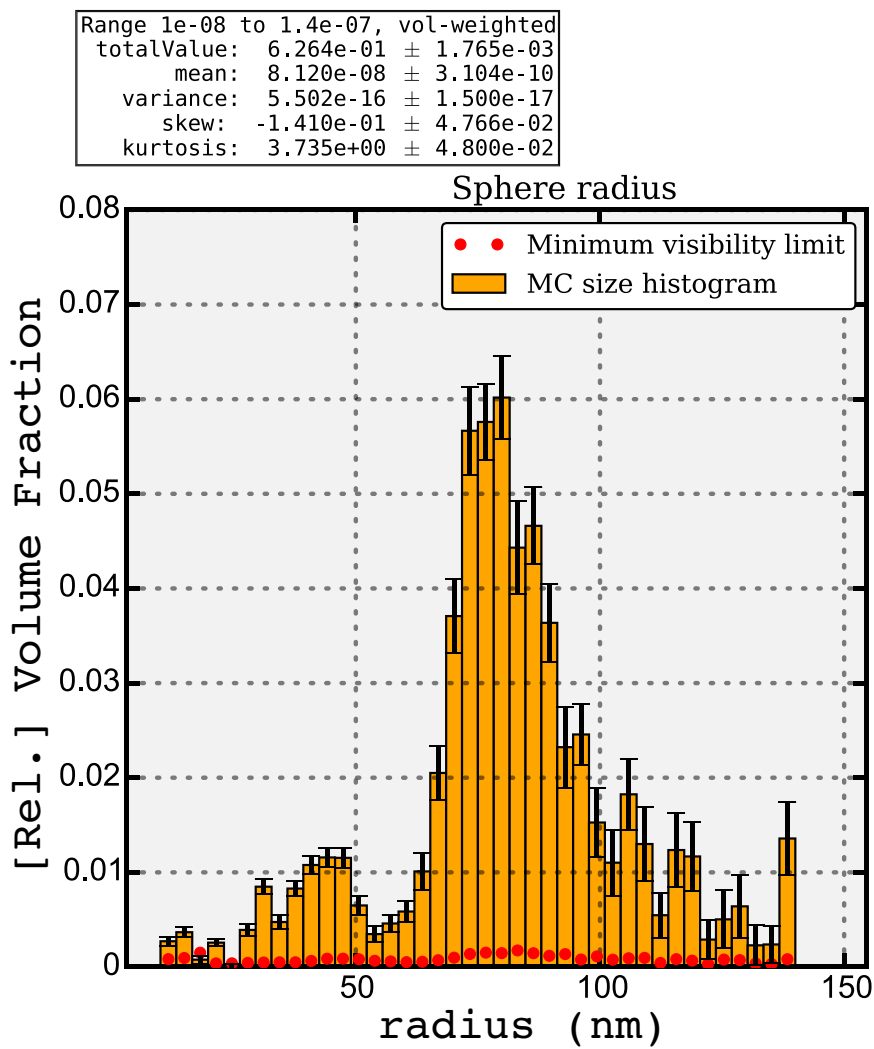}
		\caption{McSAS graphical output panel showing volume-weighted size distribution associated with the MC fit of Figure \ref{fg:denseI}.}\label{fg:denseR} 
	\end{center}
\end{figure}

A fit to within data uncertainty can be obtained using the McSAS model (c.f. Figure \ref{fg:denseI}), where the volume fraction $v_f$ is set to 0.63.
Note that the instrumental resolution has not been considered in either the classical or the MC approach. 
The main feature in the resultant size distribution (shown in Figure \ref{fg:denseR}) is indeed at the size indicated for the sample (number-weighted mean radius of 76.1(2)\,nm), but a minor component is visible at about half the radius of the main component. 

Whether these minor features are due to a breakdown of approximations in the LMA-PY structure factor, issues with sample purity or due to unaccounted-for components in the data cannot be determined without further investigation. 
What is found, however, is that a similar good fit can be obtained for other volume fractions (possible between $\approx 0.5 \leq v_f \leq 0.7$). 
Changing the volume fraction drastically affects the size distribution and demonstrates that there are a multitude of solutions accessible through adjustment of the volume fraction. 
This highlights once more that information \emph{must} be supplied on the sample to allow SAS analyses to arrive at a unique solution. 
However, the overall result is quite satisfactory and a clear improvement compared to the classical approach. 

\section{Conclusion}

A complete data analysis package is presented capable of extracting form-free size distributions from small-angle scattering patterns. 
A variety of shape functions have been included such as (core-shell) spheres, rods and ellipsoids, but also polymer models and a model for packed spheres. 
Care must be taken (in particular with more complex models) that sufficient external information is supplied to ensure a unique solution. 
When provided with data in absolute units, the resulting volume-weighted size distribution units will reflect absolute volume fractions.

\begin{ack}

The authors would like to acknowledge Peter H\o gh\o j and Xenocs for their provision of the dense spheres dataset. 
I.B. acknowledges financial support by the MIS program of BAM.

\end{ack}

%
%

\referencelist[McSASbibliography] 

\end{document}